\documentclass[aps,twocolumn,showpacs,preprintnumbers,amsmath,amssymb,nofootinbib,superscriptaddress]{revtex4-1}

\usepackage{bm}
\usepackage{footnotebackref} 

\def\mk{\mathbf{k}}

\def\BOmega{\boldsymbol{\Omega}}

\def\Bsigma{\boldsymbol{\sigma}}
\def\BSigma{\boldsymbol{\Sigma}}

\def\wsigma{\widetilde{\sigma}}

\def\Re{\mathrm{Re}}
\def\Im{\mathrm{Im}}

\newcommand{\secref}[1]{Sec.~\ref{#1}}
\newcommand{\eqnref}[1]{Eq.~(\ref{#1})}

\begin{document}

\title{Dynamical precession of spin in the two-dimensional spin-orbit coupled systems}

\author{Tsung-Wei Chen}
\email{twchen@mail.nsysu.edu.tw}
\affiliation{Department of Physics, National Sun Yat-sen University, Kaohsiung 80424, Taiwan}

\author{Zhi-Yang Huang}
\affiliation{Department of Physics, National Sun Yat-sen University, Kaohsiung 80424, Taiwan}

\author{Dah-Wei Chiou}
\email{dwchiou@gmail.com}
\affiliation{Department of Physics, National Sun Yat-sen University, Kaohsiung 80424, Taiwan}
\affiliation{Center for Condensed Matter Sciences, National Taiwan University, Taipei 10617, Taiwan}

\date{\today}

\begin{abstract}
We investigate the spin dynamics in the two-dimensional spin-orbit
coupled system subject to an in-plane ($x$-$y$ plane) constant
electric field, which is assumed to be turned on at the moment
$t=0$. The equation of spin precession in linear response to the
switch-on of the electric field is derived in terms of
Heisenberg's equation by the perturbation method up to the first
order of the electric field. The dissipative effect, which is
responsible for bringing the dynamical response to an asymptotic
result, is phenomenologically implemented \textit{\`{a} la} the
Landau-Lifshitz-Gilbert equation by introducing damping terms upon
the equation of spin dynamics. Mediated by the dissipative effect,
the resulting spin dynamics asymptotes to a stationary state,
where the spin and the momentum-dependent effective magnetic
field are aligned again and have nonzero components in the
out-of-plane ($z$) direction. In the linear response regime, the
asymptotic response obtained by the dynamical treatment is in full
agreement with the stationary response as calculated in the Kubo
formula, which is a time-independent approach treating the applied
electric field as completely time-independent. Our method provides
a new perspective on the connection between the dynamical and
stationary responses.

\end{abstract}
\pacs{71.70.Ej, 72.25.Dc, 73.43.Cd, 75.47.-m}
\maketitle

\section{Introduction}\label{sec:Intro}

The phenomenon of the spin-Hall effect is the appearance of
lateral bulk spin current in the spin-orbit coupled systems driven
solely by applying an electric field~\cite{Dya1971,Hirsch1999}.
The fact that the spin-Hall current, arising from the separation
of opposite spin orientations without breaking the time-reversal
symmetry, is dissapationless, has enormous advantages in the
development of spintronics~\cite{Pri98}. A lot of attention has
been devoted to investigating the theoretical foundations
\cite{Mura03,Sinova04,Kurebayashi14} of the spin-Hall effect and performing the
experiments \cite{Kato04} that test the validity of
theory~\cite{theo} and advance the technology of spintronics.

In two-dimensional (2D) spin-orbit coupled
systems~\cite{Rashba,Dre,Winkler02,Bulaev05}, the spin-Hall effect
becomes very important, not only for its relation to the
topological Berry phase~\cite{Xiao10,Shen2004,Chen2006,Chen2014}
but also for the development of the quantum spin-Hall
effect~\cite{Kane2005,Xiao2006,Mura2011} (2D topological
insulators) and Chern insulators~\cite{Haldane1988}, where the
definition of the bulk spin current~\cite{Shi06,Sun2005,Tan2015}
plays the key role in the bulk-edge
correspondence~\cite{Kane2005}. Recently, it was shown that the
spin-Hall effect in the two-dimensional Weyl fermion system is caused by the
spin torque current~\cite{Ong2018}. The phenomenon of the bulk
spin current is usually understood as the \emph{stationary}
response of the system to the applied in-plane electric field,
which, as well known, can be directly calculated by the Kubo
formula. However, the \emph{dynamical} origin of this response
remains rather mysterious. That is, if the applied electric field
is switched on at the moment $t=0$, how does the (expectation
value of) spin \emph{dynamically} evolve from its original
in-plane direction to yield an out-of-plane component and
eventually asymptote to the \emph{stationary} value?

The connection between the dynamical evolution and the stationary
response is essential to understanding the underlying mechanism of
the spin-Hall effect. This connection has been addressed for the
Berry curvature induced spin dynamics in the 3D p-type
semiconductor~\cite{Mura03}, the 2D $k$-linear Rashba
system~\cite{Sinova04}, the semiclassical Drude
model~\cite{Chu2007}, and the stationary response of the kinetic
equation~\cite{Mora2013}. Recently, the spin dynamics in the
honeycomb lattice~\cite{Samar2018} has also been investigated
using the Landau-Lifshitz-Gilbert equation \cite{Aharoni1996}, by
which the classical and quantum correspondence appears at
low-energy spectra. Besides theoretical importance, understanding
the dynamical evolution of spin is also crucial to the performance
of spintronic devices, which largely depends on their response
time to the applied field. In this work, we focus on the
two-dimensional spin-orbit coupled system subject to a constant
in-plane electric field switched on at $t=0$.

For the two-dimensional spin-orbit coupled systems, the
Hamiltonian before the electric field is turned on is in general
written as $H_0=\epsilon_{\mk}+\sigma_xd_x(\mk)+\sigma_yd_y(\mk)$,
where $\epsilon_{\mk}$ is the kinetic energy, $\mathbf{d}$ is
referred to as the effective magnetic field, and $\sigma_i$ are
Pauli matrices representing the real electron spin in the system.
It follows from the algebra
$[\sigma_i,\sigma_j]=2i\epsilon_{ijk}\sigma_k$ that the equation
of motion of spin (i.e., spin precession) is described by the
Larmor precession around the direction of $\mathbf{d}$. As the
quantum average of spin is parallel to the effective magnetic
field, its component in the out-of-plane ($z$) direction remains
zero if the effective magnetic field is in the in-plane ($x$-$y$)
direction. Therefore, in order to have a nonzero stationary
response of $z$ component of spin, the effective magnetic field
has to be tilted from the in-plane direction by some mechanism.

It turns out that applying an in-plane electric field provides
such a mechanism. The quantum average of spin is given by the
stationary response to the applied electric field~\cite{Chen2014},
which exhibits a nonzero component in the $z$ direction. That is,
in the presence of an in-plane electric field, the total
Hamiltonian $H=H_0+e\mathbf{E}\cdot\mathbf{x}$, when evaluated as
expectation values with respect to the eigenstates of $H_0$, can
be rendered into the form $\Bsigma\cdot\mathbf{D}(\mathbf{k})$,
where $\mathbf{D}(\mathbf{k})$ represents the new effective
magnetic field and has a nonzero $z$ component. However, the
dynamical origin of the spin-$z$ component --- especially, the
question how the electric filed tilts the effective magnetic field
from the in-plane direction --- remains obscure.

To address the dynamical issues, we treat the Hamiltonian as
\emph{time-dependent} by assuming that the electric field is
turned off for $t<0$ and turned on for $t>0$. Heisenberg's
equation is then used to solve the dynamical evolution
perturbatively up to the first order of the electric field. In
this dynamical picture, the effective magnetic field becomes
nonstatic due to the switch-on of the electric field and exhibits
a time-dependent component on the $x$-$y$ plane. The time-varying
effective magnetic field is no longer aligned with the spin for
$t>0$ and therefore drives the spin to precess around it.

The dynamical evolution is expected to asymptote to a stationary
result, where the spin and the effective magnetic field are
aligned again. To obtain the asymptotic behavior, we have to take
into account the dissipative process that attenuates and
eventually ceases the spin precession. The fundamental mechanism
for the dissipation remains unclear and complicated
\cite{Iwata1983}, but it can be phenomenologically implemented
\textit{\`{a} la} the Landau-Lifshitz-Gilbert equation
\cite{Aharoni1996} by introducing damping terms upon the equation
of spin dynamics obtained from the first-order perturbation. Via
the dissipative process, the precession of spin gives rise a
backreaction that alters the effective magnetic field and tilts it
from the $x$-$y$ plane. The resulting dynamical evolution
asymptotically approaches a stationary state, where the spin and
the effective magnetic field are aligned again and both have
nonzero components in the $z$ direction.

Meanwhile, we also directly compute the stationary response in a
time-independent approach where the electric field is treated as
always turned on. The linear term of the stationary response is
exactly equal to the asymptotic result obtained in the dynamical
analysis. Our dynamical treatment not only reveals the dynamical
origin of the spin-$z$ component in terms of the dynamical
response to the switch-on of the electric field but also
establishes its connection to the stationary response. In
particular, we uncover that the dissipative effect is crucial for
connecting the dynamical and stationary responses.

This paper is organized as follows. In \secref{sec:SpinDynamics},
the spin dynamics for spin-orbit coupled systems subject to an
constant electric field turned on at $t=0$ is derived in terms of
Heisenberg's equation up to the first order of the electric filed.
In \secref{sec:2Dsystem}, the equation of spin dynamics is
explicitly solved for a two-dimensional system. By
phenomenologically implementing the dissipative effect, the spin
dynamics is shown to approach an asymptotic result. In
\secref{sec:effmf}, we use the time-independent method to directly
calculate the stationary response of spin. The linear response of
the spin-$z$ component is exactly the same as the asymptotic
result obtained from the spin dynamics. The spin-Hall current in
relation to spin-$z$ component is also discussed in this section.
Finally, our conclusion is summarized and discussed in
\secref{sec:Conclusion}.

\section{Equation of Spin Dynamics in 3D systems}\label{sec:SpinDynamics}

In the presence of a constant and in-plane electric field
$\mathbf{E}$, the full Hamiltonian is given by
\begin{equation}\label{SD-fullH}
H=H_0+e\mathbf{E}\cdot\mathbf{x},
\end{equation}
where $H_0$ is the unperturbed Hamiltonian, and in general can be
written as the form
\begin{equation}\label{H0}
H_0=\epsilon_{\mk}+\sigma_xd_x(\mk)+\sigma_yd_y(\mk)+\sigma_zd_z(\mk),
\end{equation}
where $\sigma_x$, $\sigma_y$ and $\sigma_z$ are Pauli matrices and
represents real electron spin. The Hamiltonian $H_0$ is
time-reversal symmetric as it is invariant under the
transformation of $\sigma_i\rightarrow-\sigma_i$,
$d_{i}\rightarrow-d_i$ and $\mk\rightarrow-\mk$.

To study the spin dynamics, we have to treat the full Hamiltonian
$H$ as time-dependent with a time-dependent $\mathbf{E}$, assuming
that $\mathbf{E}(t)\rightarrow0$ as $t\rightarrow-\infty$ and
$\mathbf{E}(t)\rightarrow\mathbf{E}_0$ as $t\rightarrow\infty$.
Generally, it is challenging to solve the time-dependent
Schr\"{o}dinger (or, equivalently, Heisenberg's) equation. One
strategy is to expanded $\mathbf{E}(t)$ in a Fourier series in
time (e.g., see \cite{Mattis1958}), but adding the results of all
Fourier modes together usually only yields a formal sum
expression. Another strategy is to model $\mathbf{E}(t)$ as an
exponentially growing function given by
$\mathbf{E}(t)=\mathbf{E}_0e^{st}$ with a regularizing parameter
$s>0$ (e.g., see \cite{Koen1957,Adagideli2005}). To obtain the
static response as the asymptotic result, the regularization is
removed in the end by taking the $s\rightarrow0$ limit. However,
removing the regularization also erases the dynamical evolution in
response to the switch-on of $\mathbf{E}(t)$, which we aim to
investigate.

We adopt a different approach, assuming that $\mathbf{E}(t)$ is
switched off for $t<0$ and abruptly switched on to a constant
value $\mathbf{E}_0\equiv\mathbf{E}$ for $t>0$. More precisely,
\eqnref{SD-fullH} is modified as
\begin{equation}\label{SD-fullH'}
H=H_0+e\mathbf{E}\cdot\mathbf{x}\,\theta(t),
\end{equation}
where $\theta(t)$ is the step function. The advantage of this
modeling is that $H$ becomes time-independent again for $t>0$ and
thus the evolution operator $e^{iHt/\hbar}$ can be easily expanded
order by order for $t>0$, enabling us to conduct the perturbation
method. One might raise doubt as to whether the model is
legitimate, as the abrupt switch-on gives discontinuity at $t=0$.
It turns out nothing is illegitimate. Because the time evolution
is governed by the Schr\"{o}dinger or, equivalently, Heisenberg's
equation, which is a first-order differential equation in time,
the resulting solution remains continuous (more precisely, of
class $C^0$) with respect to $t$, even if the time-varying
potential is given discontinuous (more precisely, of class
$C^{-1}$).\footnote{For a first-order differential equation
$\alpha x'(t)+\beta x(t)+f(t)=0$, if the driving term $f(t)$ is of
class $C^{n}$, the resulting solution $x(t)$ is of class
$C^{n+1}$. Also note that the step function $\theta(t)$ is of
class $C^{-1}$.} By exploiting the fact that the solution is of
$C^0$, we focus on the dynamics for $t>0$ and take the value at
$t=0^-$ as the initial condition for
$t\geq0^+$.\footnote{Accordingly, in the rest of this paper, the
equations of motion are derived only for $t>0$ (the point $t=0^+$
is not included), unless stated otherwise.} The fact that the
resulting solution is continuous but not differentiable at $t=0$
(i.e., of class $C^0$ but not $C^1$) is just an artifact due to
the idealized (but still legitimate) modeling of abruptness of the
switch-on. In reality, the electric field can in principle be
switched on as abruptly as possible, but it remain smooth if
measured with arbitrarily high resolution in time.

According to \eqnref{SD-fullH'}, the corresponding equation of
motion of momentum is given by
\begin{equation}\label{SD-momentum}
\hbar\frac{d\mk_t}{dt}=-e\mathbf{E}\theta(t).
\end{equation}
The solution of \eqnref{SD-momentum} is
$\mk_t=\mk-e\mathbf{E}t/\hbar$ for $t\geq0$, where $\mk$ is
defined as the momentum at $t=0^-$. Because the dynamics of
momentum $\mk_t=\mk-e\mathbf{E}t/\hbar$ shows that time and
electric field couples in the form $\mathbf{E}t$, this implies
that the linear response is valid also for a very short time. The
Heisenberg picture of an observable $\mathcal{O}$ is defined as
\begin{equation}\label{SD-OH}
\mathcal{O}^H(t)=e^{iHt/\hbar}\mathcal{O}e^{-iHt/\hbar}.
\end{equation}
The dynamics of spin can always be cast in the following
Heisenberg's equation
\begin{equation}\label{SpinDynamics}
\frac{\partial}{\partial
t}\Bsigma^H(t)=\BOmega^H(t)\times\Bsigma^{H}(t)
\end{equation}
for some function $\BOmega^H(t)$, which is referred to as the
effective magnetic filed.\footnote{Rigorously speaking, it is
$-\BOmega^H$, not $+\BOmega^H$, that should be referred to as the
effective magnetic field (in $\mathbf{k}$ space). In this paper, we
nevertheless call $+\BOmega^H$ the effective magnetic field for
convenience.} \eqnref{SpinDynamics} can be exactly solved if
$\BOmega^H(t)=\BOmega$ is time-independent. For the time-dependent
effective magnetic field $\BOmega^H(t)$, as far as we know,
\eqnref{SpinDynamics} has no exact solution in general due to the
complication that the unitary transformation for diagonalizing the
effective magnetic field is time dependent.

We can solve \eqnref{SpinDynamics} perturbatively by expanding the
effective magnetic field $\BOmega^H(t)$ in series of different
orders of the applied electric field:
$\BOmega^H(t)=\BOmega_0+\BOmega_{1}(t)+o(\lambda^2)$, where
$\BOmega_0$ is of $o(\lambda^0)$ and $\BOmega_1$ is of
$o(\lambda)$. The dimensionless perturbative parameter
$\boldsymbol{\lambda}$ with $|\boldsymbol{\lambda}|<1$ is given as
a constat proportional to $\mathbf{E}$ as
\begin{equation}\label{lambda}
\boldsymbol{\lambda} = \frac{el}{\hbar\Omega_0}\mathbf{E},
\end{equation}
where $\Omega_0=2|\mathbf{d}|/\hbar$ is the interband gap of the
unperturbed system $H_0$ and $l$ denotes a length scale associated with $\mathbf{k}$.
The condition $|\boldsymbol{\lambda}|<1$ is understood as that the energy induced by $\mathbf{E}$ has to be smaller than the interband gap so that interaction between the upper and lower bands remains negligible.\footnote{Note that $\Omega_0\rightarrow0$ as $\mathbf{k}\rightarrow0$, and the condition $|\boldsymbol{\lambda}|<1$ cannot be satisfied in the vicinity of the Gamma point $\mathbf{k}=0$. It is a consequence of the fact that the perturbation method breaks down whenever the energy levels become degenerate. However, in the end when we sum up contributions over all points in the bands, only those inside the annulus circumscribed by the Fermi surface contribute. The region close to $\mathbf{k}=0$ is excluded and thus causes no problem. More rigorously, when the whole of the system is considered, \eqnref{lambda} is given by $|\boldsymbol{\lambda}| = \max_{k_F^-\leq |\mathbf{k}|\leq k_F^+}el(\mathbf{k})|\mathbf{E}|/\hbar\Omega_0(\mathbf{k})$, where $k_F^\pm$ are the upper and lower bounds of the Fermi momentum.}
The length scale $l(\mathbf{k})$ is about how far a wave packet centered at $\mathbf{k}$ travels and therefore is given by $l(\mathbf{k})=|\mathbf{v}_g(\mathbf{k})|t$ with $\mathbf{v}_g(\mathbf{k})=\boldsymbol{\nabla}_\mathbf{k}\epsilon_\mathbf{k}$ being the group velocity of the band at the point $\mathbf{k}$.
The $|\boldsymbol{\lambda}|<1$ condition then implies that the result of the
first-order perturbation is valid only if $t$ is short enough.
More precisely, the short-time condition is given by
\begin{equation}\label{ShortTime}
t < \frac{\hbar\Omega_0}{ev_g|\mathbf{E}|}.
\end{equation}

The spin is also expanded in series accordingly:
$\Bsigma^{H}(t)=\Bsigma^{H_0}(t)+\Bsigma^{\lambda}(t)+o(\lambda^2)$.
Up to the linear order of $\lambda$, \eqnref{SpinDynamics} can be
written as
\begin{equation}\label{SpinDynamics2}
\frac{\partial}{\partial
t}\Bsigma^{\lambda}(t)=\BOmega_0\times\Bsigma^{\lambda}(t)+\BOmega_1(t)\times\Bsigma^{H_0}(t),
\end{equation}
where the unperturbed equation
\begin{equation}\label{Equnper}
\frac{\partial}{\partial
t}\Bsigma^{H_0}(t)=\BOmega_0\times\Bsigma^{H_0}(t)
\end{equation}
was used. The right hand side of \eqnref{SpinDynamics2} exhibits
two different kinds of torque. The first one
$\BOmega_0\times\Bsigma^{\lambda}(t)$ gives the Larmor precession
around the direction of the static magnetic field $\BOmega_0$. The
second one $\BOmega_1(t)\times\Bsigma^{H_0}(t)$  gives a
non-Larmor precession.

As $\mathbf{E}$ is turned on at $t=0$, the initial state at $t=0$
is given by the eigenstate of the unperturbed Hamiltonian $H_0$ denoted as
$|n\mk\rangle$, with $n$ being the band index. We now apply the
expectation value $\langle n\mk|\mathcal{O}^H(t)|n\mk\rangle$,
which is also simply written as $\langle\mathcal{O}^H(t)\rangle$.
For the unperturbed system, we have $\langle
n\mk|\mathcal{O}^{H_0}(t)|n\mk\rangle=\langle
n\mk|\mathcal{O}|n\mk\rangle$. Consequently, \eqnref{SpinDynamics2} leads to
\begin{equation}\label{SpinDynamics3}
\frac{\partial}{\partial
t}\langle\Bsigma^{\lambda}(t)\rangle=\BOmega_0\times\langle\Bsigma^{\lambda}(t)\rangle+\BOmega_1(t)\times\langle\Bsigma^{H_0}(t)\rangle.
\end{equation}
\eqnref{SpinDynamics3} is the equation of motion of spin subject to the extra torque
$\BOmega_1(t)\times\langle\Bsigma^{H_0}(t)\rangle$ in addition to the Larmor torque. Furthermore,
$\langle\Bsigma^{H_0}(t)\rangle$ is parallel (or anti-parallel) to
$\BOmega_0$. This can be deduced from the fact that the left hand
side of \eqnref{Equnper} vanishes because
$\langle\Bsigma^{H_0}(t)\rangle$ is independent of time.
Therefore, the second term of \eqnref{SpinDynamics3} becomes
$\BOmega_1\times\langle\Bsigma^{H_0}(t)\rangle=\pm\BOmega_1\times\BOmega_0/\Omega_0$,
i.e., perpendicular to $\BOmega_0$. This implies that the linear
response $\langle\Bsigma^{\lambda}(t)\rangle$ is always
perpendicular to $\BOmega_0$,
\begin{equation}\label{SD-Perpen}
\BOmega_0\cdot\langle\Bsigma^{\lambda}(t)\rangle=0.
\end{equation}
\eqnref{SD-Perpen} can also be obtained as follows. Because the
first term and the second term on the right hand side of
\eqnref{SpinDynamics3} is always perpendicular to $\BOmega_0$ and
$\BOmega_0$ is time-independent, we have $\partial
(\BOmega_0\cdot\langle\Bsigma^{\lambda}(t)\rangle)/\partial t=0$.
This implies $\BOmega_0\cdot\langle\Bsigma^{\lambda}(t)\rangle$ is
a constant for all time. On the other hand, as the perturbation
term $\langle\Bsigma^{\lambda}(t)\rangle$ vanishes at $t=0$, the
constant is zero and therefore we obtain \eqnref{SD-Perpen}.
Furthermore, \eqnref{SD-Perpen} also leads to the result that
$|\langle\Bsigma^H(t)\rangle|^2=|\langle\Bsigma^{H_0}(t)\rangle|^2+o(\lambda^2)$.
This means that the magnitude of spin remains unchanged up to the
first order. That is, even though the non-Larmor torque can alter
the magnitude of spin, the change is of $o(\lambda^2)$.

Up to the linear order of the applied electric field, the second
derivative of \eqnref{SpinDynamics3} can be written as
\begin{equation}
\begin{split}
\frac{\partial^2}{\partial
t^2}\langle\Bsigma^{\lambda}(t)\rangle=&\BOmega_0\times(\BOmega_0\times\langle\Bsigma^{\lambda}(t)\rangle)\\
&+\BOmega_0\times(\BOmega_1\times\langle\Bsigma^{H_0}(t)\rangle)\\
&+\BOmega_1\times(\BOmega_0\times\langle\Bsigma^{H_0}(t)\rangle)\\
&+\frac{\partial\BOmega_1}{\partial
t}\times\langle\Bsigma^{H_0}(t)\rangle.
\end{split}
\end{equation}
For the first term, we have
\begin{equation}
\begin{split}
\BOmega_0\times(\BOmega_0\times\langle\Bsigma^{\lambda}(t)\rangle)&=\BOmega_0\cdot\langle\Bsigma^{\lambda}(t)\rangle\BOmega_0-\Omega_0^2\langle\Bsigma^{\lambda}(t)\rangle\\
&=-\Omega_0^2\langle\Bsigma^{\lambda}(t)\rangle,
\end{split}
\end{equation}
where $\BOmega_0\cdot\langle\Bsigma^{\lambda}(t)\rangle=0$ [see
\eqnref{SD-Perpen}] is used. The third term vanishes, i.e.,
$\BOmega_1\times(\BOmega_0\times\langle\Bsigma^{H_0}(t)\rangle)=0$
because $\BOmega_0$ is parallel to the unperturbed spin
$\langle\Bsigma^{H_0}(t)\rangle=\langle\Bsigma\rangle$, which is
independent of $t$. Therefore, the second derivative of
$\langle\Bsigma^{\lambda}(t)\rangle$ is given by
\begin{equation}\label{SpinDynamics4}
\begin{split}
\frac{\partial^2}{\partial
t^2}\langle\Bsigma^{\lambda}(t)\rangle=&-\Omega_0^2\langle\Bsigma^{\lambda}(t)\rangle+\BOmega_0\times(\BOmega_1\times\langle\Bsigma^{H_0}(t)\rangle)\\
&+\frac{\partial\BOmega_1}{\partial
t}\times\langle\Bsigma^{H_0}(t)\rangle.
\end{split}
\end{equation}
\eqnref{SpinDynamics3}, \eqnref{SD-Perpen} and
\eqnref{SpinDynamics4} play the key role in obtaining the spin
dynamics, and can be further simplified in two-dimensional
systems. In the next section, we will solve the spin dynamics in
the intrinsic (impurity-free) system.

\section{Intrinsic 2D systems}\label{sec:2Dsystem}

In two-dimensional systems, both $\BOmega_0$ and $\BOmega_1$ lie on
the $x$-$y$ plane; we have $\BOmega_0=(\Omega_{0x},\Omega_{0y},0)$
and $\BOmega_1=(\Omega_{1x},\Omega_{1y},0)$. This implies that
$\BOmega_1\times\BOmega_0$ is always perpendicular to the plane.
To the first order of the electric field,
$\langle\Bsigma^{\lambda}(t)\rangle$ precesses around $\BOmega_0$
but also tends to be dragged out of plane at the same time. In
this sense, we separate the solution of
$\langle\Bsigma^{\lambda}(t)\rangle$ into a harmonic term
describing the Larmor motion and an anharmonic term describing the
non-Larmor motion for the drag effect. Therefore, the first order
perturbation of spin $\langle\Bsigma^{\lambda}(t)\rangle$ can be
separated into two parts,
\begin{equation}
\langle\Bsigma^{\lambda}(t)\rangle=\BSigma^L(t)+\BSigma^N(t).
\end{equation}
$\BSigma^L$ is referred to as the Larmor component of spin and is
a harmonic function of time. $\BSigma^N$ is referred to as the
non-Larmor component of spin and is anharmonic in time.
Correspondingly, $\BSigma^L$ and $\BSigma^N$ respectively satisfy
\begin{equation}\label{EqLarmor}
\frac{\partial}{\partial t}\BSigma^L=\BOmega_0\times\BSigma^L,
\end{equation}
and
\begin{equation}\label{EqDynamical}
\frac{\partial}{\partial
t}\BSigma^N=\BOmega_0\times\BSigma^N+\BOmega_1\times\langle\Bsigma\rangle,
\end{equation}
where
\begin{equation}
\langle\Bsigma^{H_0}(t)\rangle=\langle
n\mk|\Bsigma|n\mk\rangle\equiv\langle\Bsigma\rangle
\end{equation}
was used.

\eqnref{EqLarmor} shows that $\BSigma^L$ precesses around the
static magnetic field $\BOmega_0$ harmonically and as a
consequence the magnitude $|\BSigma^L|$ is independent of time. On
the other hand, as $\BOmega_0$ and $\BOmega_1$ span the 2D plane,
the third term at the right hand side of \eqnref{SpinDynamics4} is
perpendicular to the 2D plane (the unperturbed spin
$\langle\Bsigma\rangle$ is parallel to $\BOmega_0$). Furthermore,
the second term at the right hand side of \eqnref{SpinDynamics4}
lies on the 2D plane. Therefore, the in-plane spin, denoted as
$\langle\Bsigma^{\lambda}_{\parallel}\rangle=(\langle\sigma^{\lambda}_x(t)\rangle,\langle\sigma^{\lambda}_y(t)\rangle)$.
and the out-of-plane spin, denoted as
$\langle\sigma^{\lambda}_z(t)\rangle$, satisfy the following
equations, respectively,
\begin{equation}\label{SD-xyz}
\begin{split}
&\frac{\partial^2}{\partial
t^2}\langle\Bsigma^{\lambda}_{\parallel}(t)\rangle+\Omega^2_0\langle\Bsigma^{\lambda}_{\parallel}(t)\rangle=\mathbf{G}_{\parallel}(t),\\
&\frac{\partial^2}{\partial
t^2}\langle\sigma^{\lambda}_{z}(t)\rangle+\Omega^2_0\langle\sigma^{\lambda}_{z}(t)\rangle=G_z(t),
\end{split}
\end{equation}
where
\begin{equation}\label{Spinsource}
\begin{split}
&\mathbf{G}_{\parallel}(t)=(G_x,G_y)=\BOmega_0\times\left[\BOmega_1\times\langle\Bsigma\rangle\right],\\
&G_z(t)\hat{e}_z=\frac{\partial\BOmega_1}{\partial
t}\times\langle\Bsigma\rangle.
\end{split}
\end{equation}
Importantly, $G_z$ in \eqnref{Spinsource} gives the dynamical
origin of the nonzero spin-$z$ component. We note that $G_z$ is
related to time derivative of $\BOmega_1$, which is nonzero in
general.

The unperturbed Hamiltonian of the spin-orbit coupled systems in
two dimensions can be written as ($d_z=0$)
\begin{equation}\label{SD-H0}
H_0=\epsilon_{\mk}+\sigma_xd_x(\mk)+\sigma_yd_y(\mk),
\end{equation}
where $\epsilon_{\mk}$ is the kinetic energy, and $d_x$ and $d_y$
are functions of momentum $\mk=(k_x,k_y)$, which describe the
spin-orbit coupling. The eigenenegry of \eqnref{SD-H0} satisfying
$H_0|n\mk\rangle=E_{n\mk}|n\mk\rangle$ is given by
$E_{n\mk}=\epsilon_{\mk}-nd$, where $d=\sqrt{d_x^2+d_y^2}$ and
$n=\pm$ represents the band index. For generic k-linear systems,
$d_i$ can be written as
$d_i=\sum_{j}\beta_{ij}k_j$~\cite{Chen2014}. For the k-cubic
Rahsba system~\cite{Winkler02}, $d_x=\alpha_hk^3\sin(3\phi)$ and
$d_y=-\alpha_hk^3\cos(3\phi)$. For the k-cubic Rashba-Dresselhaus
system~\cite{Bulaev05},
$d_x=[\alpha_h\sin(3\phi)+\beta_h\cos(\phi)]k^3$, and
$d_y=[-\alpha_h\sin(3\phi)+\beta_h\sin(\phi)]k^3$. The precession
frequency is given by $\Omega_0=2d/\hbar$. The expectation values
of spin with respect to the unperturbed eigenstates are given by
$\langle\Bsigma\rangle=\langle n\mk|\Bsigma|n\mk\rangle$, and we
have (see Appendix~\ref{App:ME})
\begin{equation}
\begin{split}
&\langle
n\mk|\sigma_x|n\mk\rangle=-n\frac{d_x}{d}=-n\frac{\Omega_{0x}}{\Omega_0},\\
&\langle
n\mk|\sigma_y|n\mk\rangle=-n\frac{d_y}{d}=-n\frac{\Omega_{0y}}{\Omega_0},\\
&\langle n\mk|\sigma_z|n\mk\rangle=0.
\end{split}
\end{equation}
Since
$\BOmega_0=(\Omega_{0x},\Omega_{0y},0)$ lies on the plane, the quantum average
of spin-$z$ component is zero, i.e., $\langle
n\mk|\sigma_z|n\mk\rangle=0$, as expected.
To simplify the
following calculations, we define
$d_x=d\sin\theta,~d_y=-d\cos\theta$, and we have
\begin{equation}\label{SD-partialtheta}
\begin{split}
\frac{\partial\theta}{\partial
k_a}&=\frac{\partial}{\partial k_a}\tan^{-1}\left(\frac{d_x}{-d_y}\right)\\
&=\frac{1}{d^2}\left(d_x\frac{\partial d_y}{\partial
k_a}-d_y\frac{\partial d_x}{\partial k_a}\right).
\end{split}
\end{equation}
The equation of motion
of spin for the total Hamiltonian
$H=H_0+e\mathbf{E}\cdot\mathbf{x}$ is then given by
\begin{equation}
\frac{\partial}{\partial
t}\Bsigma^H(t)=\frac{2\mathbf{d}^H(t)}{\hbar}\times\Bsigma^H(t),
\end{equation}
where $\mathbf{d}^H(t)=(d^H_x(t),d^H_y(t),0)$. The corresponding
effective magnetic field is given by
$\BOmega^H(t)=2\mathbf{d}^H(t)/\hbar$. Eq. (\ref{SD-OH}) can be
perturbatively expanded up to first order of
$eE_ax_a$~\cite{Mahan}, and the result is given by
\begin{equation}\label{SD-Exp}
\mathcal{O}^H(t)=\mathcal{O}^{H_0}(t)+eE_a[\Gamma_a,\mathcal{O}^{H_0}(t)]+o(\lambda^2),
\end{equation}
where
$\mathcal{O}^{H_0}(t)=\exp(iH_0t/\hbar)\mathcal{O}\exp(-iH_0t/\hbar)$,
and the operator $\Gamma_a$ is given by
\begin{equation}\label{SD-Gamma}
\Gamma_a=\frac{i}{\hbar}\int_0^tdt'e^{iH_0t'/\hbar}x_ae^{-iH_0t'/\hbar}.
\end{equation}
By using \eqnref{SD-Exp} and \eqnref{SD-Gamma}, we have
\begin{equation}\label{SD-d}
\begin{split}
d^H_i(t)&=e^{iHt/\hbar}d_ie^{-iHt/\hbar}\\
&=d_i-\frac{eE_a}{\hbar}\frac{\partial d_i}{\partial
k_a}t+o(\lambda^2).
\end{split}
\end{equation}
We note that the result \eqnref{SD-d} is valid in a short time as
given in \eqnref{ShortTime}. The original (i.e., zeroth-order)
effective magnetic field is given by
\begin{equation}\label{BOmega 0}
\BOmega_0=(\frac{2d_x}{\hbar},\frac{2d_y}{\hbar},0),
\end{equation}
and the first-order effective magnetic field is given by
\begin{equation}\label{BOmega 1}
\BOmega_1=\frac{2}{\hbar}(-eE_a\frac{\partial d_x}{\hbar\partial
k_a}t,-eE_a\frac{\partial
d_y}{\hbar\partial k_a}t,0)\\
\end{equation}
The magnitude of $\BOmega_0$ is $|\BOmega_0|=2d/\hbar=\Omega_0$.
We first focus on the solution of
$\langle\sigma^{\lambda}_z(t)\rangle$. The solution of the Larmor
component $\Sigma^L_z$ in \eqnref{SD-xyz} is obtained by setting
$G_z=0$, and we must have
$\Sigma^L_z=A\sin(\Omega_0t)+B\cos(\Omega_0t)$. On the other hand,
as the first-order magnetic field $\BOmega_1$ is linear in time,
its time derivative is a constant in time. The solution of the
non-Larmor component $\Sigma^N_z$ must be anharmonic, and thus, we
have $\Omega^2_0\Sigma^N_z=G_z$,
\begin{equation}\label{SigmaDz}
\begin{split}
\Sigma^N_z&=\frac{1}{\Omega^2_0}\left(\frac{\partial\BOmega_1}{\partial
t}\times\langle n\mk|\Bsigma|n\mk\rangle\right)_z\\
&=\frac{1}{\Omega^2_0}\left(\frac{\partial\Omega_{1x}}{\partial
t}\times\langle
n\mk|\sigma_y|n\mk\rangle-\frac{\partial\Omega_{1y}}{\partial
t}\times\langle n\mk|\sigma_x|n\mk\rangle\right)\\
&=-eE_a\frac{2}{\hbar^2\Omega^2_0}\left[\frac{\partial
d_x}{\partial k_a}\left(-n\frac{d_y}{d}\right)-\frac{\partial
d_y}{\partial
k_a}\left(-n\frac{d_x}{d}\right)\right]\\
&=-\frac{neE_a}{\hbar\Omega_0}\frac{\partial\theta}{\partial k_a}.
\end{split}
\end{equation}
Simply by requiring the $\Sigma^L_z+\Sigma^N_z=0$ at the initial
time $t=0$, we have the coefficient $B=-\Sigma^N_z$ and we
consequently
$\langle\sigma^{\lambda}_z(t)\rangle=\Sigma^N_z(1-\cos(\Omega_0t))+A\sin(\Omega_0t)$.
Since the $z$-component of $\BOmega_0$ is zero, i.e.,
$\Omega_{0z}=0$, the condition
$\BOmega_0\cdot\langle\Bsigma^{\lambda}(t)\rangle=0$ cannot
determine the coefficient $A$. We have to consider the time
derivative of $\langle\Bsigma^{\lambda}(t)\rangle$ together with
the solution of the in-plane component of spin.

The solution of the Larmor precession in the in-plane direction is
given by
$\BSigma_{\parallel}^L=\mathbf{C}_{\parallel}\cos(\Omega_0t)+\mathbf{K}_{\parallel}\sin(\Omega_0t)$.
The non-Larmor precession in \eqnref{SD-xyz} is given by
$\Omega^2_0\BSigma^N_{\parallel}=\mathbf{G}_{\parallel}$, and we
have
\begin{equation}\label{SigmaDxy}
\begin{split}
\BSigma^N_{\parallel}(t)&=\Sigma^N_x(t)\hat{e}_x+\Sigma^N_y(t)\hat{e}_y\\
&=\frac{1}{\Omega^2_0}\BOmega_0\times[\BOmega_1\times\langle
n\mk|\Bsigma|n\mk\rangle]\\
&=\frac{1}{\Omega^2_0}\BOmega_0\times\hat{e}_z\left(\Omega_{1x}\langle n\mk|\sigma_y|n\mk\rangle-\Omega_{1y}\langle n\mk|\sigma_x|n\mk\rangle\right)\\
&=\Sigma^N_z(\BOmega_0\times\hat{e}_z)t\\
&=\left(-d_y\frac{neE_a}{\hbar d}\frac{\partial\theta}{\partial
k_a}t\right)\hat{e}_x+\left(+d_x\frac{neE_a}{\hbar
d}\frac{\partial\theta}{\partial k_a}t\right)\hat{e}_y.
\end{split}
\end{equation}
By requiring that $\BSigma^L_{\parallel}+\BSigma^N_{\parallel}=0$
at the initial time $t=0$, we have $\mathbf{C}_{\parallel}=0$. For
the non-Larmor precession, we find that \eqnref{EqDynamical} is
satisfied by substitution of \eqnref{SigmaDz} and
\eqnref{SigmaDxy} into \eqnref{EqDynamical}. Up to this step, the
remaining coefficients are $K_x$, $K_y$ and $A$ for direct
precession. By substituting $\BSigma^L_{\parallel}$ and
$\Sigma^L_z$ into \eqnref{EqLarmor}, we can obtain $A=0$,
$K_x=-\Omega_{0y}\Sigma^N_z/\Omega_0$, and
$K_y=\Omega_{0x}\Sigma^N_z/\Omega_0$. Furthermore, by using the
condition $\BOmega_0\cdot\langle\Bsigma^{\lambda}(t)\rangle=0$, we
have $\Omega_{0x}K_x+\Omega_{0y}K_y=0$, and it is easy to check
that the result is satisfied. Therefore, the solution of $\langle
n\mk|\Bsigma^{H}(t)|n\mk\rangle$ can be written as
\begin{equation}\label{Spinsum1}
\langle
n\mk|\Bsigma^{H}(t)|n\mk\rangle=\langle\Bsigma\rangle+\BSigma^N(t)+\BSigma^L(t),
\end{equation}
where the Larmor component of spin
$\BSigma^L=(\Sigma^L_x,\Sigma^L_y,\Sigma^L_z)$ is given by
[$d_x=d\sin\theta$, $d_y=-d\cos\theta$ is used]
\begin{equation}\label{SpinL}
\begin{split}
\BSigma^L(t)=&\Sigma^N_z\cos\theta\sin(\Omega_0t)\hat{e}_x+\Sigma^N_z\sin\theta\sin(\Omega_0t)\hat{e}_y\\
&-\Sigma^N_z\cos(\Omega_0t)\hat{e}_z,
\end{split}
\end{equation}
and the non-Larmor component of spin
$\BSigma^N(t)=(\Sigma^D_x,\Sigma^D_y,\Sigma^N_z)$ is given by
\begin{equation}\label{SpinD}
\BSigma^N=\left(\BOmega_0\times\Sigma^N_z\hat{e}_z\right)t+\Sigma^N_z\hat{e}_z.
\end{equation}
For the Larmor precession [see \eqnref{SpinL}], the magnitude of
$\BSigma^L$ is $|\BSigma^L|=\Sigma^N_z$, and $\BSigma^L$ precesses
about the in-plane axis $\BOmega_0$ with frequency $\Omega_0$ as
can be verified from \eqnref{EqLarmor}. For the non-Larmor
precession, the spin-$z$ component is not necessary to be zero.
The spin-$z$ component behaves as $\Sigma^N_z(1-\cos(\Omega_0t))$,
which has a maximum value $2\Sigma^N_z$ and a minimum value $0$.

The solution of the in-plane and out-of-plane components of spin
obtained in \eqnref{Spinsum1}, \eqnref{SpinL} and \eqnref{SpinD}
can be recast into the following form [see also
\eqnref{App-Solvec}]
\begin{equation}\label{reSpin}
\begin{split}
&\langle
n\mk|\Bsigma^H_{\parallel}(t)|n\mk\rangle=-n\frac{\BOmega_0}{\Omega_0}+\Sigma^N_z\left(\BOmega_0\times\hat{e}_z\right)\left[t-\frac{\sin(\Omega_0t)}{\Omega_0}\right],\\
&\langle
n\mk|\sigma^H_{z}(t)|n\mk\rangle\hat{e}_z=\Sigma^N_z\left[1-\cos(\Omega_0t)\right]\hat{e}_z.
\end{split}
\end{equation}
The term linear in $t$ for $\langle\Bsigma^H_{\parallel}\rangle$ seems to grow arbitrarily large with $t$, but this is only an artifact due to the perturbation method, which is valid only for a short range of $t$ as delimited by \eqnref{ShortTime}. In fact, the time-dependent part of $\langle\Bsigma^H_{\parallel}\rangle$ and the whole part of $\langle\sigma^H_{z}(t)\rangle$ are both of $o(\lambda)$, reassuring $|\langle\Bsigma^H(t)\rangle|^2=1+o(\lambda^2)$.
From \eqnref{reSpin}, we see that, after $\mathbf{E}$ is turned on at $t=0$, the out-of-plane component $\langle\sigma^H_{z}(t)\rangle$ arises from zero and begins a harmonic oscillation around $\Sigma^N_z$ with the frequency $\Omega_0$.
The typical value of the frequency $\Omega_0$ on the Fermi surface
is about $10^{10}$ per second.\footnote{For pure Rashba system,
$\Omega_0\sim2\alpha k_F/\hbar$ and $k_F\sim10^{-2}\AA^{-1}$ and
$\alpha\sim10^{-3}eV\AA$ [for GaAs~\cite{Miller2003}], and we have
$\Omega_0\sim10^{10}$.}

In reality, after a while, the harmonic oscillation will decay and
$\langle\Bsigma^H(t)\rangle$ will asymptote to a new value due to
some dissipative process, which has not been taken into
consideration. The switch-on of $\mathbf{E}$ bridges the two
time-independent systems described by $H_0$ and
$H_0+e\mathbf{E}\cdot\mathbf{x}$, respectively. Measurement of the
spin in the latter system does not exhibit any harmonic
oscillation, because a time-independent system gives a stationary
solution (which will be the topic of \secref{sec:effmf}). However,
on the other hand, if the latter system is transited from the
former system, the transition will induce a harmonic oscillation
as we just calculated. Without any dissipative process, the
harmonic oscillation will remain persistent, which is unphysical
because after long enough the system should become oblivious of
when the transition has taken place.

The exact mechanism for the dissipation is unclear and
complicated, but it can be implemented phenomenologically. In
particular, the time-varying behavior of a magnetic dipole subject
to a magnetic filed can be modeled by various forms of the
Landau-Lifshitz-Gilbert equation \cite{Aharoni1996}, which
phenomenologically includes a damping term to account for the
dissipation. The Landau-Lifshitz-Gilbert equation can be obtained
more fundamentally in the context of irreversible statistical
mechanics \cite{Iwata1983}. Unfortunately, as the
Landau-Lifshitz-Gilbert equation models the time-varying behavior
only for the magnetic dipole but assumes the magnetic field as
static, it does not serve our purpose. Instead, we
phenomenologically model the dissipative process by introducing
damping terms directly upon the precessional equation
\eqnref{SD-xyz}:
\begin{equation}\label{SD-xyz damped}
\begin{split}
&\frac{\partial^2}{\partial
t^2}\langle\Bsigma^{\lambda}_{\parallel}(t)\rangle+\Omega^2_0\langle\Bsigma^{\lambda}_{\parallel}(t)\rangle
+\beta_\parallel \frac{\partial}{\partial t}\langle\Bsigma^{\lambda}_{\parallel}(t)\rangle
=\mathbf{G}_{\parallel}(t)e^{-\alpha t},\\
&\frac{\partial^2}{\partial
t^2}\langle\sigma^{\lambda}_{z}(t)\rangle+\Omega^2_0\langle\sigma^{\lambda}_{z}(t)\rangle
+\beta_z \frac{\partial}{\partial t}\langle\Bsigma^H_z(t)\rangle
=G_z(t),
\end{split}
\end{equation}
where the constants $\beta_\parallel$ and $\beta_z$ are two damping parameters responsible for damping the harmonic oscillation. For generality, we keep $\beta_\parallel$ and $\beta_z$ as two different coefficients.
Additionally, we also include an exponential decay factor $e^{-\alpha t}$ for the ``source'' term $\mathbf{G}_{\parallel}(t)$. This factor is not only prescribed to subdue the pathological trait of the linear growth in $t$ for $\mathbf{G}_{\parallel}(t)$ but in fact is required to render the damping with $\beta_\parallel$ consistent (as will be seen shortly).
Provided that the dissipative effects are strong enough, i.e., phenomenologically characterized by
\begin{equation}\label{alpha beta condition}
\frac{1}{\beta_\parallel}, \frac{1}{\beta_z}, \frac{1}{\alpha} \ll\frac{\hbar\Omega_0}{ev_d|\mathbf{E}|},
\end{equation}
the spin precession will reach an asymptotic state within the
valid range of time given by \eqnref{ShortTime}.

If we require that the solution of \eqnref{SD-xyz damped} agrees
with \eqnref{reSpin} when
$\beta_{\parallel},\beta_z,\alpha\rightarrow0$, we then have the
new solution
\begin{equation}\label{reSpin damped}
\begin{split}
&\langle n\mk|\Bsigma^H_{\parallel}(t)|n\mk\rangle
=-n\frac{\BOmega_0}{\Omega_0}
+\Sigma^N_z\frac{\Omega_0^2\left(\BOmega_0\times\hat{e}_z\right)}
{\left(\Omega_0^2-\alpha\beta_\parallel+\alpha^2\right)^2}\\
& \qquad \times
\bigg[e^{-\alpha t}(\Omega_0^2-\alpha\beta_\parallel+\alpha^2)t \\
& \qquad \qquad \mbox{} + (2\alpha-\beta_\parallel)\left(e^{-\alpha t}-e^{-\frac{\beta_\parallel t}{2}+i\tilde{\Omega}_\parallel t}\right)\\
& \qquad \qquad \mbox{}
-\Omega_0e^{-\frac{\eta\parallel t}{2}}\sin(\tilde{\Omega}_\parallel t)\bigg],
\\
&\langle n\mk|\sigma^H_{z}(t)|n\mk\rangle\hat{e}_z
=\Sigma^N_z\left[1-e^{-\frac{\beta_z t}{2}}\cos(\tilde{\Omega}_z t)\right]\hat{e}_z,
\end{split}
\end{equation}
where the new oscillatory frequencies $\tilde{\Omega}_{\parallel,z}$ are given by
\begin{equation}
\tilde{\Omega}^2_{\parallel,z}=\Omega_0^2-\left(\frac{\beta_{\parallel,z}}{2}\right)^2.
\end{equation}
Prescribing $\beta_\parallel\neq0$ gives rise to a pathological trait that $\langle\Bsigma^H_\parallel(t)\rangle$ in general becomes complex, with an imaginary part of $o(\lambda)$.
This problem can be avoided by choosing
\begin{equation}
2\alpha=\beta_\parallel.
\end{equation}
Therefore, to implement dissipation consistently, inclusion of the
damping term with $\beta_\parallel$ entails the exponential decay
term with $\alpha$.

By defining
\begin{equation}
\begin{split}
&F_0(t) = \frac{\Omega_0^3}{\Omega_0^2-\alpha\beta_\parallel+\alpha^2}e^{-\alpha t}t,\\
&F_1(t) = \frac{\Omega_0^3(2\alpha-\beta_\parallel)} {\left(\Omega_0^2-\alpha\beta_\parallel+\alpha^2\right)^2}
\left(e^{-\alpha t}-e^{-\frac{\beta_\parallel t}{2}+i\tilde{\Omega}_\parallel t}\right),\\
&F_2(t) =\frac{\Omega_0^4}{\left(\Omega_0^2-\alpha\beta_\parallel+\alpha^2\right)^2} e^{-\frac{\beta_\parallel t}{2}}\sin(\tilde{\Omega}_\parallel t),\\
&F_3(t) = e^{-\frac{\beta_z t}{2}}\cos(\tilde{\Omega}_z t),
\end{split}
\end{equation}
\eqnref{reSpin damped} can be rewritten as
\begin{equation}\label{reSpin damped'}
\begin{split}
&\langle\Bsigma^H_\parallel(t)\rangle = -n\frac{\BOmega_0}{\Omega_0}
+\Sigma^N_z\left(\frac{\BOmega_0}{\Omega_0}\times\hat{e}_z\right)\\
& \qquad\qquad\qquad\times\left[F_0(t) + F_1(t) - F_2(t)\right],\\
&\langle\sigma^H_{z}(t)\rangle\hat{e}_z = \Sigma^N_z\left[1-F_3(t)\right]\hat{e}_z,
\end{split}
\end{equation}
which follows
\begin{equation}\label{reSpin damped''}
\begin{split}
&\frac{\partial}{\partial t}\langle\Bsigma^H_\parallel(t)\rangle = \Sigma^N_z\left(\frac{\BOmega_0}{\Omega_0}\times\hat{e}_z\right)\\
&\qquad\qquad\times\left[F'_0(t) + F'_1(t) - F'_2(t)\right],\\
&\frac{\partial}{\partial t}\langle\sigma^H_{z}(t)\rangle\hat{e}_z
= \Sigma^N_z F'_3(t)\hat{e}_z.
\end{split}
\end{equation}
Eqs.~(\ref{reSpin damped'}) and (\ref{reSpin damped''}) can be organized into the following form
\begin{equation}
\frac{\partial}{\partial t}\langle\Bsigma^H(t)\rangle
= \BOmega^\lambda(t)\times \langle\Bsigma^H(t)\rangle
+o(\lambda^2),
\end{equation}
with the effective magnetic field given by
\begin{equation}\label{Omega lambda}
\begin{split}
\BOmega^\lambda(t) = & \ \ \BOmega_0\\
&\mbox{} + n\Sigma^N_z\left(\frac{\BOmega_0}{\Omega_0}\times\hat{e}_z\right)
\Big[-\Omega_0F_0(t) - \Omega_0 F_1(t)\\
&\qquad\qquad\qquad\quad \mbox{} + \big(\Omega_0F_2(t)-F'_3(t)\big)\Big]\\
&\mbox{} + n\Sigma^N_z\hat{e}_z
\bigg[-\Omega_0 + F'_0(t) + F'_1(t)\\
&\qquad\qquad\qquad\quad \mbox{} + \big(\Omega_0F_3(t)-F'_2(t)\big)\bigg]
+o(\lambda^2).
\end{split}
\end{equation}
With $\beta_{\parallel,z}=0$ and $\alpha=0$, we have $F_0(t)=\Omega_0t$, $F_1(t)=0$, $\Omega_0F_2(t)-F'_3(t)=0$ and $\Omega_0F_3(t)-F'_2(t)=0$, and consequently \eqnref{Omega lambda} leads to
\begin{equation}\label{Omega lambda trivial}
\begin{split}
\BOmega^\lambda(t)&=\BOmega_0 - n \Sigma^N_z(\BOmega_0\times\hat{e}_z) \Omega_0 t + o(\lambda)\\
&\equiv \BOmega_0 + \BOmega_1(t) + o(\lambda^2).
\end{split}
\end{equation}
This affirms the consistency that the effective magnetic field
$\BOmega^\lambda(t)$ obtained from the equation of spin precession
is identical to $\BOmega^H(t)$ in \eqnref{SpinDynamics}, where our
calculation begins.

On the other hand, with $\beta_\parallel, \beta_z, \alpha>0$, we
have $F_{0,1,2,3}(t)\rightarrow0$ for
$t\gg1/\beta_\parallel,1/\beta_z,1/\alpha$. Therefore, when $t$ is
large enough, the spin and the effective magnetic field asymptote
to the fixed values:
\begin{equation}\label{spin asym}
\langle\Bsigma^H(t)\rangle \rightarrow
-n\frac{\BOmega_0}{\Omega_0} + \Sigma^N_z\hat{e}_z + o(\lambda^2),
\end{equation}
and
\begin{equation}\label{Omega lambda asym}
\BOmega^\lambda(t)\rightarrow \BOmega_0 - n  \Omega_0
\Sigma^N_z\hat{e}_z + o(\lambda^2)
\end{equation}
In the asymptotic limit, both the spin
$\langle\Bsigma^H(t)\rangle$ and the effective magnetic field
$\BOmega^\lambda(t)$ are parallel to each other and have constant
nonzero components in the out-of-plane direction. The asymptotic
values of $\langle\Bsigma^H(t)\rangle$ and $\BOmega^\lambda(t)$
obtained from the \emph{dynamical} response to the switch-on of
the applied electric field should be the same as those obtained as
the \emph{stationary} response to the electric field that is
viewed as never-changing in time. In the next section, we will
perform the time-independent analysis upon the time-independent
Hamiltonian $H_0+e\mathbf{E}\cdot\mathbf{x}$. In terms of the
matrix elements with respect to the eigenstates of $H_0$, the full
Hamiltonian $H_0+e\mathbf{E}\cdot\mathbf{x}$ takes the form
\begin{equation}
\hat{\mathcal{H}}=\epsilon_{\mk}+\Bsigma\cdot\mathbf{D}.
\end{equation}
The direction of the effective magnetic field (i.e., $\mathbf{D}/D$)
as a stationary response is in full agreement with the asymptotic
value of the dynamical response given in \eqnref{Omega lambda
asym}.

Therefore, we have arrived at a good understanding about the
dynamical origin of the out-of-plane spin component. In the
beginning, before the electric field $\mathbf{E}$ is turned on,
the spin is aligned with the original effective magnetic filed
$\BOmega_0$, lying on the $x$-$y$ plane. When $\mathbf{E}$ is
turned on at $t=0$, it deflects $\BOmega_0$ into
$\BOmega_0+\BOmega_1+o(\lambda^2)$, which remains on the $x$-$y$
plane. The spin is no longer aligned with the new effective
magnetic field and starts to precess around it, thereby giving
rise to the spin-$z$ component. The precession of spin alters the
effective magnetic field and tilts it from the $x$-$y$ plane as a
backreaction via the dissipative process. Eventually, the
precession of spin and the evolution of the effective magnetic
field asymptotically reach a stationary balance, where the spin is
aligned again with the effective magnetic field and has a nonzero
spin-$z$ component. The dissipative effect plays a crucial role in
establishing the stationary balance.

We close this section with a remark: The mathematical result
obtained from solving \eqnref{SpinDynamics} up to the first order
of the electric field should be the same with that obtained
directly from the Heisenberg picture. The later calculation is
presented in Appendix~\ref{App:H}.

\section{Time-independent analysis}\label{sec:effmf}

As demonstrated in the previous section, an effective out-of-plane
magnetic field is dynamically generated by an in-plane electric
field and it asymptotes to an asymptotic result. In this section,
complementary to the dynamical treatment, we conduct a
time-independent analysis to directly derive the stationary
response to the applied electric field, which is now treated as
always turned on and completely time independent. It will shown
that the linear term of the stationary response is exactly equal
to the asymptotic response obtained in the dynamical treatment.

The unperturbed Hamiltonian \eqnref{SD-H0} is represented in the
spin space $|\uparrow\rangle$ and $|\downarrow\rangle$. In the
basis $|n\mk\rangle$, $H_0$ is diagonalized, while the total
Hamiltonian is not but reads as
\begin{equation}\label{Hnk}
\begin{split}
\mathcal{H}&=\left(\begin{array}{cc}
\langle +\mk|H|{+\mk}\rangle&\langle +\mk|H|{-\mk}\rangle\\
\langle -\mk|H|{+\mk}\rangle&\langle -\mk|H|{-\mk}\rangle\end{array}\right)\\
&=\left(\begin{array}{cc}
\epsilon_{\mk}-d+V_E&\langle +\mk|x_a|{-\mk}\rangle eE_a\\
\langle -\mk|x_a|{+\mk}\rangle
eE_a&\epsilon_{\mk}+d-V_E\end{array}\right),
\end{split}
\end{equation}
where we have defined $\langle+\mk|x_a|{+\mk}\rangle eE_a\equiv
V_E$ and $\langle-\mk|x_a|{-\mk}\rangle eE_a\equiv-V_E$, as in
general the vector potential $\langle n\mk|x_a|n\mk\rangle$
depends on the band index $n=\pm$. In general, the quantity
$V_E\equiv
eE_a(\langle{+\mk}|x_a|{+\mk}\rangle-\langle{-\mk}|x_a|{-\mk}\rangle)/2$
is not invariant under the gauge transformation
$|{+\mk}\rangle\rightarrow e^{i\phi_+(\mk)}|{+\mk}\rangle$,
$|{-\mk}\rangle\rightarrow e^{i\phi_-(\mk)}|{-\mk}\rangle$ unless
we choose $\phi_+(\mk)=\phi_-(\mk)$. This raises an issue of how
to define a gauge-independent spin current in the spin-Hall
effect. Following Ref.\ \cite{Shi06}, one has to apply an
intricate prescription to render the spin current gauge
independent. Nevertheless, up to the first order of the electric
field, the linear response is independent of $V_E$ and thus free
of this problem as will be seen shortly.

For the off-diagonal matrix elements of $x_a$, we can write $\langle
\pm\mk|x_a|{\mp\mk}\rangle$ in terms of matrix elements of
$\sigma_z$, which is valid for all choice of wave functions, as
proved in Appendix~\ref{App:ME}. Using \eqnref{App-xnmk}, we have
\begin{equation}\label{xnk}
\begin{split}
&\langle +\mk|x_a|{-\mk}\rangle=\frac{1}{2}\langle
+\mk|\sigma_z|{-\mk}\rangle\frac{\partial\theta}{\partial
k_a},\\
&\langle -\mk|x_a|{+\mk}\rangle=\frac{1}{2}\langle
-\mk|\sigma_z|{+\mk}\rangle\frac{\partial\theta}{\partial k_a}.
\end{split}
\end{equation}
Substituting \eqnref{xnk} into \eqnref{Hnk}, $\mathcal{H}$ can be
written as
\begin{equation}\label{Hnk2}
\mathcal{H}=\epsilon_{\mk}+a_x\tau_x+a_y\tau_y+a_z\tau_z,
\end{equation}
where
\begin{equation}
\begin{split}
&a_x=\frac{1}{2}\Re\langle-\mk|\sigma_z|{+\mk}\rangle\frac{\partial\theta}{\partial
k_a}eE_a\\
&a_y=\frac{1}{2}\Im\langle-\mk|\sigma_z|{+\mk}\rangle\frac{\partial\theta}{\partial
k_a}eE_a\\
&a_z=V_E-d.
\end{split}
\end{equation}
The matrices $\tau_i$, which are called pseudo-spin matrices,  are
mathematically Pauli matrices, but they are not real spin. This
can also be seen as follows. The position operator $\mathbf{x}$ and
the electric field $\mathbf{E}$ are even under the time-reversal transformation.
The real spin is odd $\sigma_i\rightarrow-\sigma_i$ and the momentum
is also odd under the time-reversal operation $\mk\rightarrow-\mk$.
The effective magnetic field $\mathbf{d}$ must be odd under
the time-reversal operation $d_x\rightarrow-d_x$ and
$d_y\rightarrow-d_y$. This implies that the Hamiltonian $H_0$ is
invariant under the time-reversal transformation. Since the Hamiltonian
$H_0$ is invariant under time-reversal transformation, the
Hamiltonian in the basis $|n\mk\rangle$ must not break the time
reversal symmetry. Therefore, we find that $a_x$, $a_y$ and
$a_z$ are even under time reversal operation, and thus $\tau_i$
must be even, which means that
$\tau_i$ are not the real spin. In order to transform \eqnref{Hnk2}
back to the real spin, we have to transform the spin in basis
$|n\mk\rangle$ to the spin space $|\uparrow\rangle$ and
$|\downarrow\rangle$.

The original spin $\sigma_i$ in basis $|n\mk\rangle$ can be
written in terms of the new spin matrices denoted as $\wsigma_i$.
For the spin-$z$ component, we have
\begin{equation}\label{tsigmaz}
\begin{split}
\wsigma_z&\equiv\left(\begin{array}{cc}
\langle +\mk|\sigma_z|{+\mk}\rangle&\langle +\mk|\sigma_z|{-\mk}\rangle\\
\langle -\mk|\sigma_z|{+\mk}\rangle&\langle
-\mk|\sigma_z|{-\mk}\rangle\end{array}\right)\\
&=\frac{2}{\frac{\partial\theta}{\partial
k_a}eE_a}\left(a_x\tau_x+a_y\tau_y\right).
\end{split}
\end{equation}
For spin-$x$ nd spin-$y$ components, we have
\begin{equation}\label{tsigmax}
\begin{split}
\wsigma_x&\equiv\left(\begin{array}{cc}
\langle +\mk|\sigma_x|{+\mk}\rangle&\langle +\mk|\sigma_x|{-\mk}\rangle\\
\langle -\mk|\sigma_x|{+\mk}\rangle&\langle
-\mk|\sigma_x|{-\mk}\rangle\end{array}\right)\\
&=-\frac{d_x}{d}\tau_z+\frac{d_y}{d}\frac{2}{\frac{\partial\theta}{\partial
k_a}eE_a}\left(a_y\tau_x-a_x\tau_y\right).
\end{split}
\end{equation}
and
\begin{equation}\label{tsigmay}
\begin{split}
\wsigma_y&\equiv\left(\begin{array}{cc}
\langle +\mk|\sigma_y|{+\mk}\rangle&\langle +\mk|\sigma_y|{-\mk}\rangle\\
\langle -\mk|\sigma_y|{+\mk}\rangle&\langle
-\mk|\sigma_y|{-\mk}\rangle\end{array}\right)\\
&=-\frac{d_y}{d}\tau_z-\frac{d_x}{d}\frac{2}{\frac{\partial\theta}{\partial
k_a}eE_a}\left(a_y\tau_x-a_x\tau_y\right).
\end{split}
\end{equation}
By using \eqnref{App-offsigmaz}, it is easy to show that
$\wsigma_i$ satisfies the algebra of the Pauli matrices,
i.e., $\{\wsigma_i,\wsigma_j\}=2\delta_{ij}$
and
$[{\wsigma_i},{\wsigma_j}]=2i\epsilon_{ijk}{\wsigma_k}$.
By using \eqnref{tsigmaz}, \eqnref{tsigmax} and \eqnref{tsigmay},
the Hamiltonian \eqnref{Hnk2} can be written in terms of
$\widetilde{\sigma_i}$ and the result is given by
\begin{equation}\label{Hnk3}
\mathcal{H}=\epsilon_{\mk}+D_x\wsigma_x+D_y\wsigma_y+D_z\wsigma_z,
\end{equation}
where
\begin{equation}\label{Dxyz}
\begin{split}
&D_x=-\frac{d_x}{d}a_z=d_x-\frac{d_x}{d}V_E,\\
&D_y=-\frac{d_y}{d}a_z=d_y-\frac{d_y}{d}V_E,\\
&D_z=\frac{1}{2}\frac{\partial\theta}{\partial k_a}eE_a.
\end{split}
\end{equation}
Therefore, the Hamiltonian in terms of the expectation values with
respect to $|{\pm\mk}\rangle$ can be again cast into the form
$\mathcal{H}=\epsilon_{\mk}+\mathbf{D}\cdot\widetilde{\Bsigma}$.
That is, the spin is aligned with an effective magnetic field
given by \eqnref{Dxyz}. We also find that
$\mathbf{D}=(D_x,D_y,D_z)$ are odd under time-reversal operation.
Therefore, the Hamiltonian in terms of the expectation values with
respect to $|\pm\mk\rangle$ can be cast into the form
\begin{equation}\label{Heff}
\hat{\mathcal{H}}=\epsilon_{\mk}+D_x\sigma_x+D_y\sigma_y+D_z\sigma_z,
\end{equation}
where $\sigma_i$ are pauli matrices. Importantly,
\eqnref{Heff} shows that the $z$-component of effective magnetic
field is non-zero. The effective magnetic field is being tilted up
in the presence of an electric field. In the absence of electric
field, \eqnref{Heff} goes back to the unperturbed Hamiltonian
$H_0=\epsilon_{\mk}+\sigma_xd_x+\sigma_yd_y$. It should be noted
that $D_x$ and $D_y$ in \eqnref{Dxyz} contain unphysical
gauge-dependent pieces involving $V_E$. Nevertheless, the
gauge-dependent terms are of $o(\lambda)$ and are exactly
cancelled out when we compute $\mathbf{D}/D$ (i.e., the direction
of $\mathbf{D}$) in the linear response regime. Noting
that $D^2=d^2-2V_Ed+o(\lambda^2)$, we have
\begin{equation}\label{Dexpand}
\frac{\mathbf{D}}{D}=\frac{d_x}{d}\hat{e}_x+\frac{d_y}{d}\hat{e}_y+\frac{1}{2d}\frac{\partial\theta}{\partial
k_a}eE_a\hat{e}_z+o(\lambda^2).
\end{equation}
Therefore, up to $o(\lambda^2)$, the result of \eqnref{Dexpand} is
gauge independent. We find that the direction of the effective
magnetic field given in \eqnref{Dexpand} is exactly the same as that of
\eqnref{Omega lambda asym}.

The eigenstates of $\hat{\mathcal{H}}$ are given by
\begin{equation}\label{eigenstateH}
\begin{split}
&|\Psi_{-\mk}\rangle=\frac{1}{\sqrt{2(1+\hat{D}_z)}}\left(\begin{array}{c}
1+\hat{D}_z\\
\hat{D}_x+i\hat{D}_y\end{array}\right),\\
&|\Psi_{+\mk}\rangle=\frac{1}{\sqrt{2(1+\hat{D}_z)}}\left(\begin{array}{c}
-\hat{D}_x+i\hat{D}_y\\
1+\hat{D}_z\end{array}\right),
\end{split}
\end{equation}
where $\hat{D}_i=D_i/D$ and $D=\sqrt{D_x^2+D_y^2+D_z^2}$. The
corresponding eigenenergies are given by
$\hat{\mathcal{H}}|\Psi_{\ell\mk}\rangle=\mathcal{E}_{\ell\mk}|\Psi_{\ell\mk}\rangle$
with $\mathcal{E}_{\ell\mk}=\epsilon_{\mk}-\ell D$. In the absence
of electric field, the eigenenergy is given by
$\mathcal{E}_{\ell\mk}=\epsilon_{\mk}-\ell d$, which is the
eigenenergy of the unperturbed Hamiltonian $H_0$. As
mentioned above, the spin is aligned in the direction
$\mathbf{D}/D$, and by using \eqnref{eigenstateH}, we have
$\langle\Psi_{\ell\mk}|\Bsigma|\Psi_{\ell\mk}\rangle=-\ell\mathbf{D}/D$.
In the presence of the electric field, up to the first order of
$\mathbf{E}$, we have
\begin{equation}\label{spin expectation}
\begin{split}
\langle\Psi_{\ell\mk}|\Bsigma|\Psi_{\ell\mk}\rangle&=-\ell\mathbf{D}/D\\
&=-\ell\frac{d_x}{d}\hat{e}_x-\ell\frac{d_y}{d}\hat{e}_y,-\frac{\ell}{2d}\frac{\partial\theta}{\partial
k_a}eE_a\hat{e}_z+o(\lambda^2).
\end{split}
\end{equation}
On the other hand, consider the projection of spin on the
$\mathbf{D}$ direction, which is given by
\begin{equation}\label{ConSpin}
\widehat{\BSigma}_{c}=\frac{1}{D^2}\mathbf{D}(\Bsigma\cdot\mathbf{D}).
\end{equation}
By noting that $D^2=d^2-2V_Ed+o(\lambda^2)$, the spin
$\widehat{\BSigma}_{c}=(\widehat{\Sigma}_{cx},\widehat{\Sigma}_{cy},\widehat{\Sigma}_{cz})$
in the unperturbed basis $|n\mk\rangle$ is given by
\begin{equation}\label{wsigma}
\begin{split}
&\langle n\mk|\widehat{\Sigma}_{cx}|n\mk\rangle=-n\frac{d_x}{d}+o(\lambda^2),\\
&\langle n\mk|\widehat{\Sigma}_{cy}|n\mk\rangle=-n\frac{d_y}{d}+o(\lambda^2),\\
&\langle
n\mk|\widehat{\Sigma}_{cz}|n\mk\rangle=-n\frac{1}{2d}\frac{\partial\theta}{\partial
k_a}eE_a+o(\lambda^2).
\end{split}
\end{equation}
It can be shown that
$\langle\Psi_{n\mk}|\Bsigma|\Psi_{n\mk}\rangle=\langle
n\mk|\widehat{\BSigma}_{c}|n\mk\rangle+o(\lambda^2)$. Importantly,
compare to \eqnref{SigmaDz}, we have
\begin{equation}\label{mainresult}
\langle n\mk|\widehat{\Sigma}_{cz}|n\mk\rangle=\Sigma^N_z.
\end{equation}
\eqnref{wsigma} and \eqnref{mainresult} are in full agreement with the
asymptotic response obtained in \eqnref{Omega lambda asym}.

We close this section by showing the relation between the spin
current and $\Sigma^N_z$ by using the Kubo formula~\cite{Mahan}.
The spin-$z$ component satisfies the following continuity
equation~\cite{Shi06,Chen2009}
\begin{equation}\label{Conti}
\frac{\partial}{\partial t}\Psi^{\dag}
S_z\Psi+\nabla_i\Re\left[\Psi^{\dag}J^z_i\Psi\right]=\Re\left[\Psi^{\dag}\frac{dS_z}{dt}\Psi\right],
\end{equation}
where the conventional spin current $J^{z}_i$ is given by
\begin{equation}
\begin{split}
J^{z}_i&=\frac{1}{2}\{J\hbar\sigma_z,v_i\}\\
&=J\frac{\partial\epsilon_{\mk}}{\partial k_i}\sigma_z,
\end{split}
\end{equation}
where $v_i=\partial H/\hbar\partial k_i$ and  $J=1/2$ for spin
$1/2$ and so on. The Kubo formula for spin current~\cite{Mura03}
is given by
\begin{equation}\label{Kubo1}
\mathcal{J}^z_i=\frac{2q\hbar}{V}\sum_{n(\neq
n')}f_{n\mk}\frac{\Im\langle n\mk|J^z_i|n'\mk\rangle\langle
n'\mk|v_j|n\mk\rangle}{(E_{n\mk}-E_{n'\mk})^2}E_j,
\end{equation}
where $n=\pm$, $q=-e$ for an electron and $f_{n\mk}$ is the
Fermi-Dirac distribution. Use the matrix elements
\eqnref{App-offsigmaz} and \eqnref{App-v} shown in Appendix
\ref{App:ME}, we have
\begin{equation}\label{Kubo2}
\langle n\mk|\sigma_z|n'\mk\rangle\langle n'\mk|v_j|n\mk\rangle\\
=\frac{-2id^2}{\hbar(E_{n\mk}-E_{n'\mk})}\frac{\partial\theta}{\partial
k_j},
\end{equation}
which is purely imaginary. Substituting \eqnref{Kubo2} into
\eqnref{Kubo1}, we have
\begin{equation}\label{Kubo3}
\begin{split}
\mathcal{J}^z_i&=\frac{J}{V}\sum_{n\mk}f_{n\mk}\frac{\partial\epsilon_{\mk}}{\partial
k_i}\left(\frac{-n}{2d}\right)\frac{\partial\theta}{\partial k_j}eE_j\\
&=\frac{J}{V}\sum_{n\mk}f_{n\mk}\frac{\partial\epsilon_{\mk}}{\partial
k_i}\Sigma^N_z\\
&=\frac{J}{V}\sum_{n\mk}f_{n\mk}\frac{\partial\epsilon_{\mk}}{\partial
k_i}\langle n\mk|\widehat{\Sigma}_{cz}|n\mk\rangle+o(\lambda^2).
\end{split}
\end{equation}
In the second equality of \eqnref{Kubo2}, the definition of
$\Sigma^N_z$ [see \eqnref{SigmaDz}] was used. The third equality
of \eqnref{Kubo2} can be directly obtained from \eqnref{Heff} in
the linear response regime, and the result is in agreement with
the Kubo formula. Furthermore, since the source term in
\eqnref{Conti} is $d\langle
n\mk|\sigma^{H}_z(t)|n\mk\rangle/dt\rightarrow d\Sigma^N_z/dt=0$,
and thus we find that the spin current shown in \eqnref{Kubo3} is
the conserved spin current. Another definition of spin current
from the source term $J^{\tau}_i=\frac{1}{2}\{x_i,d\sigma_z/dt\}$
(spin-torque current) has been shown to be zero in the presence of
constant electric field~\cite{Chen2006,Chen2014}, which is in
agreement with the present result. The spin-torque current is
non-zero only when the electric field is nonhomogeneous in the
space $\mathbf{E}\exp(iqx)$, and $J^{\tau}_i$ will be the rate of
change of the torque spin density with respective to $q$ in the
limit $q\rightarrow0$~\cite{Shi06}. In addition, \eqnref{wsigma}
shows that the linear response of the in-plane spin is zero. This
can be seen as follows. By using \eqnref{App-offsigmax} and
\eqnref{App-v}, we have
\begin{equation}\label{zerox}
\begin{split}
&\langle n\mk|\sigma_x|n'\mk\rangle\langle
n'\mk|v_j|n\mk\rangle\\
&=\frac{2id_y}{E_{n'\mk}-E_{n\mk}}\frac{-2id^2}{\hbar(E_{n\mk}-E_{n'\mk})}|\langle
n\mk|\sigma_z|n'\mk\rangle|^2\frac{\partial\theta}{\partial k_j}\\
&=\frac{-4d_yd^2}{(E_{n\mk}-E_{n'\mk})^2}|\langle
n\mk|\sigma_z|n'\mk\rangle|^2\frac{\partial\theta}{\partial k_j},
\end{split}
\end{equation}
which is purely real, and therefore the imaginary part is zero. For
$\sigma_y$, similar to the derivation shown in \eqnref{zerox} (by
using \eqnref{App-sigmay} and \eqnref{App-v}), it can be shown
that $\langle n\mk|\sigma_y|n'\mk\rangle\langle
n'\mk|v_j|n\mk\rangle$ is also purely real.

\section{Conclusion}\label{sec:Conclusion}
We obtain the dynamical equation of spin in two-dimensional
spin-orbit coupled systems by solving Heisengerg's equation
perturbatively up to the linear order of the applied electric
field, which is assumed to be turned on at $t=0$. As shown in
Eqs.~(\ref{BOmega 0}) and (\ref{BOmega 1}), the switch-on of the
electric field deflects the effective magnetic field from its
original direction by giving a time-dependent component on the
$x$-$y$ plane. As the spin is no longer aligned with the effective
magnetic field, it starts to precess around the new direction.

Taking into account the dissipative effect that attenuates and
eventually ceases the spin precession, we phenomenologically add
damping terms upon the equation of spin dynamics as in
\eqnref{SD-xyz damped}. The solution of the resulting dynamics is
given in \eqnref{reSpin damped} for the spin and \eqnref{Omega
lambda} for the effective magnetic field. When $t$ is large
enough, the dynamical solution asymptotes to an asymptotic state
given by Eqs.~(\ref{spin asym}) and (\ref{Omega lambda asym}),
where the spin and the effective magnetic field are aligned again
and exhibit nonzero components in the $z$ direction.

On the other hand, treating the applied electric field as always
turned on, we also directly compute the stationary response in the
time-independent approach by projecting the full Hamiltonian on
the spin space as in \eqnref{Heff}. The stationary response is
obtained in Eqs.~(\ref{wsigma}) [and (\ref{mainresult})], which is
exactly equal to the asymptotic result \eqref{spin asym} obtained
from the dynamical treatment. The direction of effective magnetic
field [\eqnref{Dexpand}] is also in agreement with that of the
asymptotic result [\eqnref{Omega lambda asym}]. Furthermore, the
relation between the stationary response of the effective magnetic
field and the spin current is derived, and the result is in
agreement with the Kubo formula.

Our dynamical treatment reveals the dynamical origin of the
spin-$z$ component and provides a method to study the connection
between the dynamical and stationary responses. The dissipative
effect is found to be crucial for establishing the connection.
However, our prescription of dissipative effect remains
phenomenological and it should be derived more fundamentally by
the methods of irreversible statistical mechanics following the
lines of Ref.~\cite{Iwata1983}. Furthermore, it has been shown in
Ref.~\cite{Morawetz2015} that, in a many-body interacting system,
the stationary effective magnetic field is created by the sum of
the vector $\mathbf{d}$ in \eqnref{H0} that accounts for the
spin-orbit coupling and a mean-field contribution derived from the
many-body problem. This seems to suggest that the dissipative
effect might be understood as the result of many-body interaction
in a mean-field theory approach. In the dynamical treatment, we
study the evolution of the expectation values of the kind $\langle
n\mk|\cdot|n\mk\rangle$ but disregard the off-diagonal terms of
the kind $\langle {+\mk}|\cdot|{-\mk}\rangle$ and
$\langle{-\mk}|\cdot|{+\mk}\rangle$, while in the time-independent
approach, both are included [see \eqnref{tsigmay}]. The fact that
the dynamical treatment with dissipation asymptotically leads to
the stationary result of the time-independent approach strongly
suggests that the dissipative effect in the dynamical picture is
closely related to the equilibrium of the interband transition in
the stationary picture. This relation should become more
transparent if the dissipation can be more fundamentally derived.

\begin{acknowledgments}
T.-W.~Chen would like to thank Wang-Chuang Kuo for valuable discussions on the stationary response.
This work was supported in part by the Ministry of Science and Technology, Taiwan under the Grant MOST 106-2112-M-110-010.
\end{acknowledgments}

\appendix

\section{Matrix Elements}\label{App:ME}
In this appendix, we calculate unperturbed matrix elements of the
spin and velocity operators used in this article without
specifying any form of the wave functions $|n\mk\rangle$. By using
$\{\sigma_x,H_0\}=\{\sigma_x,\epsilon_k+\sigma_xd_x+\sigma_yd_y\}=2\epsilon_{\mk}\sigma_x+2d_x$,
and $\langle n\mk|\{\sigma_x,H_0\}|n\mk\rangle=2\langle
n\mk|\sigma_x|n\mk\rangle E_{n\mk}=2\epsilon_{\mk}\langle
n\mk|\sigma_x|n\mk\rangle-2nd\langle n\mk|\sigma_x|n\mk\rangle$.
We have the diagonal matrix element of $\sigma_x$ in the helicity
basis,
\begin{equation}\label{App-dsigmax}
\langle n\mk|\sigma_x|n\mk\rangle=-n\frac{d_x}{d}.
\end{equation}
For the off-diagonal matrix elements, we note that
$[\sigma_x,H_0]=[\sigma_x,\epsilon_k+\sigma_xd_x+\sigma_yd_y]=2i\sigma_z
d_y$. It follows
\begin{equation}\label{App-offsigmax}
\langle
n\mk|\sigma_x|m\mk\rangle=\frac{2id_y}{E_{m\mk}-E_{n\mk}}\langle
n\mk|\sigma_z|m\mk\rangle.
\end{equation}
Similar to the derivation, for the spin-$y$ component, we have
\begin{equation}\label{App-sigmay}
\begin{split}
\langle n\mk|\sigma_y|n\mk\rangle&=-n\frac{d_y}{d},\\
\langle
n\mk|\sigma_y|m\mk\rangle&=\frac{-2id_x}{E_{m\mk}-E_{n\mk}}\langle
n\mk|\sigma_z|m\mk\rangle.
\end{split}
\end{equation}
For the spin-$z$ component, we have
$\{\sigma_z,H_0\}=2\epsilon_{\mk}\sigma_z$, and this implies
$-2nd\langle n\mk|\sigma_z|n\mk\rangle=0$. If the splitting $d$ is
nonzero (the spin-orbit coupling does not vanish and $\mk\neq0$),
we have
\begin{equation}\label{App-dsigmaz}
\langle n\mk|\sigma_z|n\mk\rangle=0.
\end{equation}
The off-diagonal matrix element of $\sigma_z$ cannot be further
determined, and in general $\langle n\mk|\sigma_z|m\mk\rangle$
depends on the choice of wave functions $|n\mk\rangle$. However,
from $\sigma_z^2=1$, we have $\langle
{+\mk}|\sigma_z\sigma_z|{+\mk}\rangle=1$. By inserting
$\sum_{n}|n\mk\rangle\langle n\mk|=1$ into the result, we obtain
$\langle{+\mk}|\sigma_z|{+\mk}\rangle\langle
{+\mk}|\sigma_z|{+\mk}\rangle+\langle
{+\mk}|\sigma_z|{-\mk}\rangle\langle{-\mk}|\sigma_z|{+\mk}\rangle=1$.
Because of $\langle{+\mk}|\sigma_z|{+\mk}\rangle=0$ as shown
above, we have in general
\begin{equation}\label{App-offsigmaz}
|\langle n\mk|\sigma_z|m\mk\rangle|^2=1,~n\neq m.
\end{equation}

The velocity operator is defined as $v_b=\partial
H_0/\hbar\partial k_b=\partial\epsilon_{\mk}/\hbar\partial
k_b+\sigma_x(\partial d_x/\hbar\partial k_b)+\sigma_y(\partial
d_y/\hbar\partial k_b)$. By using Eqs.~(\ref{App-offsigmax}),
(\ref{App-sigmay}), the off-diagonal matrix element of velocity
operator is given by
\begin{equation}\label{App-v}
\langle
n\mk|v_b|m\mk\rangle=\frac{-2id^2}{\hbar(E_{m\mk}-E_{n\mk})}\langle
n\mk|\sigma_z|m\mk\rangle\frac{\partial\theta}{\partial k_b},
\end{equation}
where \eqnref{SD-partialtheta} was used. The off-diagonal matrix
element of $v_b$ is related to the position operator $x_b$ by
$v_b=dx_b/dt=[x_b,H_0]/i\hbar$. By using $\langle
n\mk|m\mk\rangle=0$ for $n\neq m$, we have
\begin{equation}\label{App-xnmk}
\langle
n\mk|x_b|m\mk\rangle=\frac{i\hbar}{E_{m\mk}-E_{n\mk}}\langle
n\mk|v_b|m\mk\rangle.
\end{equation}
For the diagonal part of $v_b$, by using Eqs. (\ref{App-dsigmax})
and (\ref{App-sigmay}), we can obtain $\langle
n\mk|v_b|n\mk\rangle=\partial E_{n\mk}/\hbar\partial k_b$.

\section{Heisebgerg operator}\label{App:H}
In this appendix, we directly obtain the solutions of spin
dynamics in terms of the time-evolving spin operators in the
Heisenburg picture. We will show that the result obtained from the
method of solving equation of motion is the same with that
obtained from the Heisenberg time evolution method. The
Hamiltonian under consideration is given by
\begin{equation}
H_0=\epsilon_{\mk}+K,
\end{equation}
where $K\equiv\sigma_xd_x+\sigma_yd_y$, and $d_x$ and $d_y$
depends on the momentum and the spin-orbit coupling. The band
index $n$ is defined as $E_{n\mk}=\epsilon_{\mk}-nd$, and the
diagonal matrix element of $K$ is $\langle
n\mk|K|n\mk\rangle=-nd$. The Pauli matrices $\sigma_x$, $\sigma_y$
and $\sigma_z$ satisfy $\{\sigma_i,\sigma_j\}=2\delta_{ij}$ and
$[\sigma_i,\sigma_j]=2i\epsilon_{ijk}\sigma_k$. By using
$K^{2}=d_x^2+d_y^2\equiv d^2$, we have $K^{2m+1}=d^{2m}K$. The
time evolution operator of the unperturbed Hamiltonian is given by
\begin{equation}\label{expH0}
e^{iH_0t/\hbar}=e^{i\epsilon_{\mk}t/\hbar}\left[\cos\left(\frac{d}{\hbar}t\right)+\frac{iK}{d}\sin\left(\frac{d}{\hbar}t\right)\right].
\end{equation}
The time evolution of $\sigma_x$, $\sigma_y$ and $\sigma_z$ under
the unperturbed Hamiltonian can be written as
\begin{equation}
\begin{split}
\Bsigma^{H_0}(t)=&e^{iH_0t/\hbar}\Bsigma e^{-iH_0t/\hbar}\\
=&\Bsigma\cos^2\left(\frac{d}{\hbar}t\right)+\frac{i}{2d}\sin\left(\frac{2d}{\hbar}t\right)[K,\Bsigma]\\
&+\frac{1}{d^2}\sin^2\left(\frac{d}{\hbar}t\right)\left(K\Bsigma
K\right).
\end{split}
\end{equation}
The operator $[K,\Bsigma]$ is given by
\begin{equation}
\begin{split}
&[K,\sigma_x]=-2i\sigma_zd_y,~[K,\sigma_y]=2i\sigma_zd_x,\\
&[K,\sigma_z]=-2i(d_x\sigma_y-d_y\sigma_x).
\end{split}
\end{equation}
The operator $K\Bsigma K$ is given by
\begin{equation}
\begin{split}
&K\sigma_xK=2d_y(d_x\sigma_y-d_y\sigma_x)+\sigma_xd^2,\\
&K\sigma_yK=-2d_x(d_x\sigma_y-d_y\sigma_x)+\sigma_yd^2,\\
&K\sigma_zK=-\sigma_zd^2.
\end{split}
\end{equation}
Therefore, we have
\begin{equation}\label{App-sigmaxyz}
\begin{split}
&\sigma_x^{H_0}(t)=\sigma_x+\frac{d_y}{d}\sin\left(\Omega_0t\right)\sigma_z\\
&\qquad\qquad \mbox{}-\frac{d_y(\mathbf{d}\times\Bsigma)_z}{d^2}\left[\cos\left(\Omega_0t\right)-1\right]\\
&\sigma_y^{H_0}(t)=\sigma_y-\frac{d_x}{d}\sin\left(\Omega_0t\right)\sigma_z\\
&\qquad\qquad\mbox{} +\frac{d_x(\mathbf{d}\times\Bsigma)_z}{d^2}\left[\cos\left(\Omega_0t\right)-1\right]\\
&\sigma_z^{H_0}(t)=\sigma_z\cos\left(\Omega_0t\right)+\frac{(\mathbf{d}\times\Bsigma)_z}{d}\sin\left(\Omega_0t\right),\\
\end{split}
\end{equation}
where $\Omega_0=2d/\hbar$ was used. In order to simplify the
calculations, we define $d_x=d\sin\theta$ and $d_y=-d\cos\theta$,
and we have \eqnref{SD-partialtheta}. For position operator $x_a$,
we have
\begin{equation}
\begin{split}
x_a^{H_0}(t)&=e^{iH_0t/\hbar}x_ae^{-iH_0t/\hbar}\\
&=x_a+e^{iH_0t/\hbar}\left(i\frac{\partial}{\partial
k_a}e^{-iH_0t/\hbar}\right),
\end{split}
\end{equation}
By substitution of \eqnref{expH0}, we have

\begin{equation}\label{App-x}
\begin{split}
x_a^{H_0}(t)=&x_a+\left[\frac{\partial\epsilon_{\mk}}{\hbar\partial
k_a}+\frac{K}{2d}\frac{\partial\Omega_0}{\partial
k_a}\right]t+\frac{1}{2}\sigma_z\frac{\partial\theta}{\partial
k_a}\left[\cos(\Omega_0t)-1\right]\\
&+\frac{1}{2d}\frac{\partial\theta}{\partial
k_a}\left(\mathbf{d}\times\Bsigma\right)_z\sin(\Omega_0t).
\end{split}
\end{equation}

In the presence of applied electric field, the operator
$\mathcal{O}$ can be perturbatively expanded up to the first order of
the electric field~\cite{Mahan}, and the result is given by
\begin{equation}\label{App-Exp}
\mathcal{O}^H(t)=\mathcal{O}^{H_0}(t)+eE_a[\Gamma_a,\mathcal{O}^{H_0}(t)]+o(\lambda^2),
\end{equation}
where the operator $\Gamma_a$ is given by
\begin{equation}\label{App-Gamma}
\Gamma_a=\frac{i}{\hbar}\int_0^tdt'e^{iH_0t'/\hbar}x_ae^{-iH_0t'/\hbar}.
\end{equation}
By substituting \eqnref{App-sigmaxyz} and \eqnref{App-x} into
\eqnref{App-Exp} and \eqnref{App-Gamma}, after straightforward
calculations, we have
\begin{equation}
\begin{split}
\Bsigma^H(t)=&\Bsigma^{H_0}(t)+\frac{i}{\hbar}[\mathbf{A}_{1}\sigma_x+\mathbf{A}_{2}\sigma_y\\
&+\mathbf{A}_{3}\sigma_z+\mathbf{B}K+\mathbf{C}\left(\mathbf{d}\times\Bsigma\right)_z],
\end{split}
\end{equation}
where $A_{x1}$, $A_{x2}$, $A_{x3}$, $B_x$ and $C_x$ are for the spin-$x$
component, the spin-$y$ component and so on. The separation is convenient for obtaining
the diagonal matrix element $\langle n\mk|(\cdots)|n\mk\rangle$
because $\langle n\mk|\sigma_z|n\mk\rangle=0$ and $\langle
n\mk|(\mathbf{d}\times\Bsigma)_z|n\mk\rangle=0$. The results of
the spin-$z$ component are given by
\begin{equation}
\begin{split}
A_{x1}=&it\frac{d_y}{d^2}[\cos(\Omega_0)-1]\frac{\partial d_y}{\partial k_a},\\
A_{x2}=&-it\frac{d_y}{d^2}[\cos(\Omega_0)-1]\frac{\partial d_x}{\partial k_a}+i\frac{\partial\theta}{\partial k_a}\left[\frac{1}{\Omega_0}\sin(\Omega_0t)-t\right],\\
A_{x3}=&it\frac{\partial}{\partial k_a}\left(\frac{d_y}{d}\sin(\Omega_0t)\right)-\frac{it^2}{2d}\frac{\partial\Omega_0}{\partial k_a}d_y\cos(\Omega_0t)\\
&+\frac{i\hbar}{2d^2}\frac{\partial\theta}{\partial k_a}[\cos(\Omega_0t)-1]d_x,\\
B_{x}=&i\frac{\partial\theta}{\partial k_a}\left[\frac{1}{\Omega_0}\sin(\Omega_0t)-t\right]\frac{d_y}{d^2}[\cos(\Omega_0t)-1]\\
&-\frac{i\hbar}{2d^2}\frac{\partial\theta}{\partial k_a}[\cos(\Omega_0t)-1]\frac{d_y}{d}\sin(\Omega_0t),\\
C_x=&-\frac{it^2}{2d}\frac{\partial\Omega_0}{\partial
k_a}\frac{d_y}{d}\sin(\Omega_0t)-it\frac{\partial}{\partial
k_a}\left[\frac{d_y}{d^2}(\cos(\Omega_0t)-1)\right].
\end{split}
\end{equation}
By using $\langle n\mk|\sigma^{H_0}_x(t)|n\mk\rangle=\langle
n\mk|\sigma_x|n\mk\rangle=-nd_x/d$, $\langle
n\mk|\sigma_y|n\mk\rangle=-nd_y/d$, $\langle
n\mk|K|n\mk\rangle=-nd$ and neglecting the irrelevant terms $C_x$
and $A_{x3}$, we obtain
\begin{equation}
\begin{split}
\langle
&n\mk|\sigma^H_x(t)|n\mk\rangle\\
=&-n\frac{d_x}{d}+\frac{i}{\hbar}\bigg\{\frac{d_y}{d^2}\left[\cos(\Omega_0t)-1\right]\frac{int}{d}\left(\frac{\partial\mathbf{d}}{\partial
k_a}\times\mathbf{d}\right)_z\\
&-in\frac{\partial\theta}{\partial
k_a}\left[\frac{\hbar}{2d}\sin(\Omega_0t)-t\right]\frac{d_y}{d}\cos(\Omega_0t)\\
&\qquad\mbox{}+\frac{in\hbar}{2d^2}\frac{\partial\theta}{\partial
k_a}[\cos(\Omega_0t)-1]d_y\sin(\Omega_0t)\bigg\}\\
=&-n\frac{d_x}{d}+\frac{neE_a}{\hbar}\frac{d_y}{d}\frac{\partial\theta}{\partial
k_a}\left[\frac{1}{\Omega_0}\sin(\Omega_0t)-t\right].
\end{split}
\end{equation}
Similar to the derivation of $\langle
n\mk|\sigma^{H}_x(t)|n\mk\rangle$, the spin-$y$ component is given
by
\begin{equation}
\langle
n\mk|\sigma^{H}_x(t)|n\mk\rangle=-n\frac{d_y}{d}-\frac{neE_a}{\hbar}\frac{d_x}{d}\frac{\partial\theta}{\partial
k_a}\left(\frac{1}{\Omega_0}\sin(\Omega_0t)-t\right).
\end{equation}
For the spin-$z$ component, the coefficients are given by
\begin{equation}
\begin{split}
&A_{z1}=-it\sin(\Omega_0t)\frac{1}{d}\frac{\partial d_y}{\partial k_a},\\
&A_{z2}=it\sin(\Omega_0t)\frac{1}{d}\frac{\partial d_x}{\partial k_a},\\
&A_{z3}=-\frac{it^2}{2}\frac{\partial\Omega_0}{\partial k_a}\sin(\Omega_0t),\\
&B_{z}=\frac{it}{d}\frac{\partial\theta}{\partial k_a}\sin(\Omega_0t)+\frac{i\hbar}{2d^2}\frac{\partial\theta}{\partial k_a}[\cos(\Omega_0t)-1],\\
&C_{z}=\frac{it^2}{2d}\frac{\partial\Omega_0}{\partial
k_a}\cos(\Omega_0t)-it\sin(\Omega_0t)\frac{1}{d^2}\frac{\partial
d}{\partial k_a}.
\end{split}
\end{equation}
By using $\langle n\mk|\sigma^{H_0}_x(t)|n\mk\rangle=\langle
n\mk|\sigma_x|n\mk\rangle=-nd_x/d$, $\langle
n\mk|\sigma_y|n\mk\rangle=-nd_y/d$, $\langle
n\mk|K|n\mk\rangle=-nd$, $\langle n\mk|\sigma_z|n\mk\rangle=0$ and
neglecting the irrelevant terms $C_x$ and $A_{x3}$, we can obtain
\begin{equation}
\langle
n\mk|\sigma_z^{H}(t)|n\mk\rangle=\frac{neE_a}{\hbar\Omega_0}\frac{\partial\theta}{\partial
k_a}[\cos(\Omega_0t)-1].
\end{equation}
By using the definition of $\Sigma^N_z$, the components of spin
can be written as
\begin{equation}\label{App-Solvec}
\begin{split}
&\langle
n\mk|\Bsigma^H_{\parallel}(t)|n\mk\rangle=-n\frac{\BOmega_0}{\Omega_0}+\Sigma^N_z\left(\BOmega_0\times\hat{e}_z\right)\left[t-\frac{\sin(\Omega_0t)}{\Omega_0}\right],\\
&\langle
n\mk|\sigma^H_{z}(t)|n\mk\rangle\hat{e}_z=\Sigma^N_z\left[1-\cos(\Omega_0t)\right]\hat{e}_z.
\end{split}
\end{equation}
and we can see that \eqnref{App-Solvec} is exactly the same with
\eqnref{Spinsum1}.

\end{document}